\documentclass{llncs}
\usepackage{amssymb}
\usepackage{amsmath}
\usepackage{framed}
\usepackage{csquotes}

\usepackage{url}
\usepackage{algpseudocode}
\usepackage{algorithm}
\usepackage{graphicx}
\usepackage{epstopdf}
\usepackage{enumitem}

\usepackage{comment}

\newcommand{\bwh}{{\sc BWS-$H$} }
\newcommand{\bwhp}{{\sc BWS-$H$ Problem} }
\newcommand{\hfb}{{\sc $H$-Free Bipartitioning} }
\newcommand{\hfbp}{{\sc $H$-Free Bipartitioning Problem} }
\newcommand{\hfqc}{{\sc $H$-Free $q$-Coloring} }

\urldef{\mailsa}\path|{aravind, subruk, cs14resch01002}@iith.ac.in|

\begin{document}

\title{Bipartitioning Problems on Graphs with Bounded Tree-Width}

\author{N. R. Aravind \and Subrahmanyam Kalyanasundaram \and Anjeneya Swami Kare}
%\thanks{Author is a faculty member of University of Hyderabad.}
\institute{Department of Computer Science and Engineering,\\IIT Hyderabad, Hyderabad, India\\ \mailsa} %\and
%University of Hyderabad, Hyderabad, India}
\maketitle

\begin{abstract}
For an undirected graph $G$, we consider the following
%bipartitioning
problems: given a fixed graph
$H$, can we partition the vertices of $G$ into two non-empty sets $A$ and $B$
such that neither the induced
graph $G[A]$ nor $G[B]$ contain $H$ (i) as a subgraph? (ii) as an induced subgraph?
These problems are NP-complete and are expressible in monadic second order logic (MSOL).
The MSOL  formulation, together with Courcelle's theorem implies linear time solvability
on graphs with bounded tree-width. This approach yields algorithms with running time
%$f(\phi, t)n^{O(1)}$,
$f(|\varphi|, t)\cdot n$,
where $|\varphi|$ is the length of the MSOL formula, $t$ is the tree-width of the
graph and $n$ is the number of vertices of the graph. The dependency of $f(|\varphi|, t)$
on $|\varphi|$ can be as bad as a tower of exponentials.

In this paper, we present explicit combinatorial algorithms for these problems
for graphs $G$ whose tree-width is bounded. We obtain $2^{O(t^r)}\cdot n$ time algorithms when $H$
is any fixed graph of order $r$. In the special case when $H = K_r$, a complete graph on $r$ vertices,
we get an $2^{O(t+r \log t)}\cdot n$ time algorithm.

The techniques can be extended to provide FPT algorithms to determine the smallest number $q$
such that $V$ can be partitioned into $q$ parts such that none of the parts have $H$ as a subgraph
(induced subgraph).
\end{abstract}

%\keywords{Bipartitioning, Bi-coloring}

\section{Introduction}
\label{intro}
Let $G = (V, E)$ be an undirected graph on $n$ vertices. In the classical $k$-coloring problem,
we need to color the vertices of the graph using at most $k$ colors such that no pair of adjacent vertices
are of the same color. The $k$-coloring problem is NP-complete for $k \geq 3$ and
this problem, and its variants, have been studied extensively under various settings.
%from the researchers across the world.
For $k = 2$, this is equivalent to testing whether
the graph is bipartite or not, which is of course solvable in polynomial time.

We consider the following generalization of the 2-coloring problem: we need to $2$-color the vertices of the
graph such that the subgraphs induced by the respective color classes do not have a fixed graph $H$
as a subgraph\footnote{The classical 2-coloring problem is obtained by setting $H = K_2$.}.
We call this problem the {\sc Bipartitioning without Subgraph $H$ Problem} or {\sc BWS-$H$ Problem} in short.
\begin{framed}
\vspace*{-0.10cm}
\noindent \bwhp\\
\textbf{Instance:} An undirected graph $G = (V, E)$. \\
\textbf{Question:} Can $V$ be partitioned into two non-empty sets $A, B$ such that neither of the induced graphs $G[A]$ and $G[B]$
have $H$ as a subgraph?
\vspace*{-0.10cm}
\end{framed}
%
%In other words,  given a fixed graph $H$, we want to partition the vertices of the
%graph $G$ into two sets $A$ and $B$ such that both the induced subgraphs $G[A]$ and $G[B]$ do not have $H$
%as a subgraph.
 We also study the variant of the problem where $H$ does not appear as an induced subgraph. We call this the
{\sc $H$-Free Bipartitioning Problem}.

%\noindent Note that the property we are looking for is \emph{not having $H$ as a subgraph} and not restricted to
%only induced subgraphs.

\begin{framed}
\vspace*{-0.10cm}
\noindent \hfbp\\
\textbf{Instance:} An undirected graph $G = (V, E)$. \\
\textbf{Question:} Can $V$ be partitioned into two non-empty sets $A, B$ such that neither of the induced graphs $G[A]$ and $G[B]$
have $H$ as an induced subgraph?
\vspace*{-0.10cm}
\end{framed}

The \bwh problem is NP-complete~\cite{farrugia2004} unless $H = K_2$. Recently, Karpi\'{n}ski~\cite{Karpinski2017}
gave an alternate proof for the NP-completeness of the problem when $H = C_r$, a cycle of fixed length $r$.
The \hfbp is NP-complete~\cite{Achlioptas1997} as long as $H$ has $3$ or more vertices.
For fixed $H$, both these problems can be expressed in monadic second order logic (MSOL).
The well-known Courcelle's theorem~\cite{courcelle1990,courcelle1992} states that any graph property that is
expressible in MSOL is solvable
in linear time for graphs with bounded tree-width.
The resulting algorithms have a
 running time $f(|\varphi|, t)\cdot n$, where $|\varphi|$ is the length of the MSOL formula and
$t$ is the tree-width of the graph. Even though the algorithms run in linear time,
the dependency of $f$ on $|\varphi|$ and $t$ can be quite bad. Indeed in the worst case
$f(|\varphi|, t)$ can be a tower of exponentials. Considering this, it is preferable
to have explicit combinatorial algorithms, since such algorithms are more efficient
and are amenable to a precise running time analysis.

In this paper, we give combinatorial algorithms for both \bwh and \hfb problems. Our main result is 
the following:
\begin{theorem}
There are $2^{O(t^r)} \cdot n$ time algorithms that solves the \bwh  and \hfb problems for any arbitrary fixed
$H$ ($|V(H)| = r$), on graphs with tree-width at most $t$.
\end{theorem}
We also obtain a much faster 
$2^{O(t+r \log t)}\cdot n$ time algorithm when $H = K_r$, a complete graph on $r$ vertices. 
 Note that in this case, the \bwh problem and \hfb problem coincide.
%\begin{itemize}
% \item $2^{O(t+r \log t)}\cdot n$ time algorithm when $H = K_r$, a complete graph on $r$ vertices. 
% Note that in this case, the \bwh problem and \hfb problem coincide.
% \item $2^{O(t^2)}\cdot n$ time algorithm when $H = C_4$, a cycle of length $4$.
% \item $2^{O(t^r)}\cdot n$ time algorithm when $H$ is any fixed graph of order $r$.
%\end{itemize}

Graph bipartitioning with other constraints have been  explored in the past.
The degree bounded bipartitioning problem asks to partition the vertices of $G$ into two sets $A$ and $B$
such that the maximum degree in the induced subgraphs $G[A]$ and $G[B]$ are at most $a$ and $b$ respectively.
Xiao and Nagamochi~\cite{xiao2017} proved that this problem is NP-complete for any non-negative integers $a$ and $b$
except for the case $a = b = 0$, in which case the problem is equivalent to testing whether $G$ is bipartite.
%They also studied the problem from parameterized complexity perspective.
Other variants that place
constraints on the degree of the vertices within the partitions have also been studied~\cite{Cowen1986,Bazgan}.
Wu, Yuan and Zhao~\cite{wu1996} showed the NP-completeness of the variant that asks to partition the vertices of the
graph $G$ into two sets such that both the induced graphs are acyclic.
A generalization of the \hfb problem called \hfqc has been mentioned in~\cite{rao2007}.
%They showed that this problem is NP-complete for graphs of maximum degree 5 and
%polynomial-time solvable for graphs of maximum degree at most 4.

Farrugia~\cite{farrugia2004} showed the NP-completeness
of a general variant of the problems called $(\mathcal{P},\mathcal{Q})$-coloring problem.
Here, $\mathcal{P}$ and $\mathcal{Q}$ are any additive induced-hereditary graph properties.
The problem asks to partition the vertices of $G$ into $A$ and $B$ such that $G[A]$ and $G[B]$
have properties $\mathcal{P}$ and $\mathcal{Q}$
respectively. %Farrugia proved that this problem is NP-complete except for the case of bipartite testing.

\begin{comment}
Chv\'{a}tal~\cite{Chavtal1984} studied another NP-Hard bipartitioning problem called Matching Cut problem.
a matching cut is a partition of vertices into two sets such that the edges across the sets induce a matching.
The matching cut problem is the problem of deciding whether a given graph has a matching cut. This problem
is also studied from parameterized complexity perspective~\cite{Bonsma2009,kratsch2016,mccocoa}.
\end{comment}

\section{Preliminaries}
We write $f(n) = O^*(g(n))$ if $f(n) = O(g(n) n^c)$ for some constant $c > 0$. Let $G = (V, E)$ be an undirected graph.
For $u \in V$, the set of all neighbors of $u$ (\emph{open neighborhood}) is denoted by $N(u)$. The \emph{closed neighborhood}
of $u$, denoted by $N[u]$, is defined as $N[u] = N(u) \cup \{u\}$. For a vertex set $S \subseteq V$,
the subgraph induced by $S$ is denoted by $G[S]$.
%For a vertex set $S \subseteq V$, $G \backslash S$ denotes the graph $G[V \backslash S]$.
When there is no ambiguity,
we use the simpler notations $S \backslash x$ to denote $S \backslash \{x\}$ and $S \cup x$ to denote $S \cup \{x\}$.
We denote the set of all $k$ sized subsets of the set $S$ by $\binom{S}{k}$. We use $uv$ to denote the edge $\{ u, v\}$
for convenience. We follow the standard graph theoretic terminology from~\cite{Diestel2005}.

A parameterized problem is a language $L \subseteq \Sigma^* \times \mathbb{N}$, where $\Sigma$ is a fixed and finite
alphabet. For $(x,k) \in \Sigma^* \times \mathbb{N}$, $k$ is referred to as the parameter.
A parameterized problem $L$ is \emph{fixed parameter tractable (FPT)} if there is an algorithm $A$, a computable
non-decreasing function $f:\mathbb{N} \rightarrow \mathbb{N}$ and a constant $c$ such that, given
$(x,k) \in \Sigma^* \times \mathbb{N}$ the algorithm $A$ correctly decides whether $(x,k) \in L$ in time
bounded by $f(k).|x|^c$. For more details on parameterized algorithms refer to~\cite{fpt-book}.

\label{prelims}
A \emph{tree decomposition} of $G$ is a pair $(T, \{ X_i, i \in I\})$, where for $i \in I$, $X_i \subseteq V$ (usually called bags) and $T$
is a tree with elements of $I$ as the nodes such that:
\begin{enumerate}
  \item For  each vertex $v \in V$, there is an $i \in I$ such that $v \in X_i$.
  \item For each edge $\{u,v\}\in E$, there is an $i \in I$ such that $\{u, v\} \subseteq X_i$.
  \item For each vertex $v \in V$, $T[\{ i \in I | v \in X_i\}]$ is connected.
\end{enumerate}
The width of the tree decomposition is $\max_{i \in I} (|X_i| - 1)$. The tree-width of $G$
is the minimum width taken over all tree decompositions of $G$ and we denote it as $t$. For more details on tree-width,
we refer the reader to~\cite{twbw1991}.
A rooted tree decomposition is called a \emph{nice tree decomposition},
%Kloks~\cite{Kloks1994} introduced the notion of \emph{nice tree decomposition},
%which is a rooted tree decomposition where
if every node $i \in I$ is one of the following types:
\begin{enumerate}
  \item Leaf node: For a leaf node $i$, $X_i = \emptyset$.
  \item Introduce Node: An introduce node $i$ has exactly one child $j$ and there is a vertex $v \in V\backslash X_j$
such that $X_i = X_j \cup \{ v\}$.
%($v \notin X_j$).
  \item Forget Node: A forget node $i$ has exactly one child $j$ and there is a vertex $v \in V\backslash X_i$
such that $X_j = X_i \cup \{ v\}$.
  \item Join Node: A join node $i$ has exactly two children $j_1$ and $j_2$ such that $X_i = X_{j_{1}} = X_{j_{2}}$.
\end{enumerate}
The notion of \emph{nice tree decomposition} was introduced by Kloks~\cite{Kloks1994}.
Every graph $G$ has a nice tree decomposition with $ |I| = O(n)$ nodes and width equal to the tree-width of $G$.
Moreover, such a decomposition can be found in linear time if the tree-width is bounded.

\subsection{Overview of the Techniques Used}
In the rest of the paper, we assume that the nice tree decomposition is given.
Let $i$ be a node in the nice tree decomposition, $X_i$ is the bag of vertices associated with the node $i$.
Let $T_i$ be the subtree rooted at the node $i$, $G[T_i]$ denote the graph induced by all the vertices in $T_i$.

We use dynamic programming on the nice tree decomposition to solve the problems for different $H$. We
process the nodes of nice tree decomposition according to its post order traversal.
We say that a partition $A, B$ of $G$ is a \emph{valid} partition if neither $G[A]$ nor
$G[B]$ have $H$ as a subgraph. At each node
$i$, we check each bipartition $(A_i, B_i)$ of the bag $X_i$ to see if $(A_i, B_i)$ leads to a valid
partition in the graph $G[T_i]$. For each partition, we also keep some extra information that
will help us to detect if the partition leads to an invalid partition at some ancestral
(parent) node.
We have four types of nodes in the tree decomposition -- leaf, introduce, forget and join nodes.
In the algorithm, we explain the procedure for updating the information at each
of these above types of nodes and consequently, to certify whether a partition is valid or not.

In Section~\ref{algokr}, we discuss algorithm for the case $H=K_r$, a complete graph on $r$ vertices.
In Section~\ref{algoc4}, we discuss algorithm for the \bwh problem when $H=C_4$, a cycle of length $4$.
In Section~\ref{algoH}, the algorithm for the \bwh problem for a fixed arbitrary graph $H$ is presented.
Presenting algorithms for $H=K_r$ and $H=C_4$ initially will help in the exposition, as they
will help to understand the setup before moving to the more involved generalized case.
Finally, we explain how the algorithm for the \hfb problem can be obtained by modifying the 
algorithm for the \bwh problem in Section~\ref{sec:hfree}.

\section{Bipartitioning without $K_r$}
\label{algokr}
We consider the \bwh problem when $H = K_r$, a complete graph on $r$ vertices.

Let $\Psi = (A_i, B_i)$ be a partition of a bag $X_i$. We set $M_i[\Psi]$ to $1$ if there exist a partition
$(A, B)$ of $V[T_i]$ such that $A_i \subseteq A$, $B_i \subseteq B$ and both $G[A]$ and $G[B]$ are $K_{r}$-free.
Otherwise, $M_i[\Psi]$ is set to $0$.

\vspace{0.1in}
\noindent\textbf{Leaf node:}
For a leaf node $\Psi = (\emptyset, \emptyset)$ and $M_{i}[\Psi] = 1$.

\vspace{0.1in}
\noindent \textbf{Introduce node:}
Let $j$ be the only child of the node $i$. Suppose, $v \in X_i$ is the new vertex present in $X_i$, $v \notin X_j$.
Let $\Psi = (A_i, B_i)$ be a partition of $X_i$. If $G[A_i]$ or $G[B_i]$ has $K_r$ as a subgraph, we set $M_i[\Psi]$
to $0$. Otherwise, we use the following cases to compute $M_i[\Psi]$ value. Since $v$ cannot have forgotten neighbors,
it can form a $K_r$ only within the
bag $X_i$.
\begin{description}
  \item[Case 1:] $v \in A_i$, $M_i[\Psi] = M_j[\Psi']$, where $\Psi' = (A_i \backslash v, B_i)$.
  \item[Case 2:] $v \in B_i$, $M_i[\Psi] = M_j[\Psi']$, where $\Psi' = (A_i, B_i \backslash v)$.
\end{description}

\noindent \textbf{Forget node:}
Let $j$ be the only child of the node $i$. Suppose, $v \in X_j$ is the vertex missing in $X_i$,
$v \notin X_i$. Let $\Psi = (A_i, B_i)$ be a partition of $X_i$. If $G[A_i]$ or $G[B_i]$ has $K_r$ as a subgraph,
we set $M_i[\Psi]$ to $0$. Otherwise, $M_i[\Psi] = \max\{ M_{j}[\Psi'], M_{j}[\Psi''] \}$, where, $\Psi' = (A_i \cup v, B_i)$
and $\Psi'' = (A_i, B_i \cup v)$.

\vspace{0.1in}
\noindent \textbf{Join node:}
Let $j_1$ and $j_2$ be the children of the node $i$. $X_i = X_{j_1} = X_{j_2}$ and $V(T_{j_1}) \cap V(T_{j_2}) = X_i$.
%There are no edges between $V(T_{j_1}) \backslash X_i$ and $V(T_{j_2}) \backslash X_i$.
Let $\Psi = (A_i, B_i)$ be a
partition of $X_i$. If $G[A_i]$ or $G[B_i]$ has $K_r$ as a subgraph, we set $M_i[\Psi]$ to $0$. Otherwise, we use the
following expression to compute $M_i[\Psi]$ value.
Since there are no edges between $V(T_{j_1}) \backslash X_i$ and $V(T_{j_2}) \backslash X_i$,
a $K_r$ cannot
contain forgotten vertices from both $T_{j_1}$ and $T_{j_2}$.

\begin{equation}\label{kreq1}
\nonumber
  M_i[\Psi] =
  \left\{
    \begin{aligned}[lr]
      &1, & \hbox{If $M_{j_{1}}[\Psi] = 1$ and $M_{j_{2}}[\Psi] = 1$.} \\
      &0, & \hbox{Otherwise.}
    \end{aligned}
  \right.
\end{equation}

Correctness of the algorithm implied from the correctness of $M_i[\Psi]$ values,
which can be proved using bottom up induction on nice tree decomposition.
$G$ has a valid bipartitioning if there exists a $\Psi$ such that
$M_r[\Psi] = 1$, where $r$ is the root node of the nice tree decomposition.
The total time complexity of the algorithm
is $2^tt^r n = O^*(2^{t+r\log t})$. With this we state the following theorem.

\begin{theorem}
\label{kr:thm1}
There is an $O(2^{t+r\log t}n)$ time algorithm  that solves the \bwh problem when $H = K_r$,
on graphs with tree-width at most $t$.
\end{theorem}

\section{Bipartitioning without $C_4$}
\label{algoc4}
In this section, we describe the combinatorial algorithm for the \bwh problem for the case
when $H = C_4$, a cycle of length $4$. As stated, the problem can be expressed in MSOL.
An MSOL formulation of the \bwh problem for the case  $H = C_4$ is given below.

\begin{align*}
  \exists V_1 \subseteq V: \exists V_2 \subseteq V: (V_1  \cap V_2 = \emptyset)  \wedge  (V_1 \cup V_2 = V)  \wedge  \neg(V_1 = \emptyset) \wedge \neg(V_2 = \emptyset) \wedge \\
  \neg(\exists u_1 \in V_1:\exists u_2 \in V_1:\exists u_3 \in V_1:\exists u_4 \in V_1:\\
   (u_1u_2 \in E)\wedge (u_2u_3 \in E)\wedge (u_3u_4 \in E)\wedge (u_4u_1 \in E)) \wedge\\
   \neg(\exists u_1 \in V_2:\exists u_2 \in V_2:\exists u_3 \in V_2:\exists u_4 \in V_2:\\
   (u_1u_2 \in E)\wedge (u_2u_3 \in E)\wedge (u_3u_4 \in E)\wedge (u_4u_1 \in E)).\\
\end{align*}

The predicates $V_1  \cap V_2 = \emptyset$, $V_1  \cup V_2 = V$ and $V_1 = \emptyset$ can be rewritten as follows:
\begin{align*}
  V_1  \cap V_2 = \emptyset \Longleftrightarrow \neg \exists v \in V : v \in V_1 \wedge v \in V_2,\\
  V_1  \cup V_2 = V  \Longleftrightarrow \forall v \in V: v \in V_1 \vee v \in V_2,\\
  V_1 = \emptyset \Longleftrightarrow \forall v \in V: \neg(v \in V_1).
\end{align*}

Note that a cycle of length $4$ is formed when a pair of (adjacent or non-adjacent) vertices have
two or more common neighbors. If a graph has no $C_4$ then any vertex pair can have at most one common neighbor.
Let $X_i$ be a bag at the node $i$ of the nice tree decomposition. We guess a partition $(A_i, B_i)$ of the bag
$X_i$.
For each pair of vertices from $A_i$ (similarly $B_i$), we also guess if the pair has exactly one common forgotten
neighbor in part $A$ (similarly $B$) of the partition.
%in the same side of the partition as that of the pairs.
We check if the above guesses lead to a valid partitioning in the subgraph $G[T_i]$, which is the graph induced
by the vertices in the node $i$ and all its descendent nodes. Below we formally explain the technique.

Let $\Psi = (A_i, B_i, P_i, Q_i)$ be a $4$-tuple defined as follows: $(A_i, B_i)$ is a partition of $X_i$,
$P_i \subseteq \binom{A_i}{2}$ and $Q_i \subseteq \binom{B_i}{2}$. Intuitively, $P_i$ and $Q_i$ are the set of those
pairs that have exactly one  common forgotten neighbor.

We define $M_i[\Psi]$ to be $1$ if there is a partition $(A, B)$ of $V(T_i)$ such that:
\begin{enumerate}%[label=(\roman*)]
\item $A_i \subseteq A$ and $B_i \subseteq B$.

\item Every pair in $P_i$ has exactly one common neighbor in $A \backslash A_i$.

\item Every pair in $\binom{A_i}{2} \setminus P_i$ does not have a common neighbor in $A \backslash A_i$.

\item Every pair in $Q_i$ has exactly one common neighbor in $B \backslash B_i$.

\item Every pair in $\binom{B_i}{2} \setminus Q_i$ does not have a common neighbor in $B \backslash B_i$.

\item $G[A]$ and $G[B]$ do not have $C_4$ as a subgraph.
\end{enumerate}
Otherwise, $M_i[\Psi]$ is set to $0$. Suppose there exists a 4-tuple $\Psi$ such that
$M_r[\Psi] = 1$, where $r$ is the root of the nice tree decomposition.
Then the above conditions 1 and 6 ensure that $G$ can be partitioned in the required manner.

%We say that a 4-tuple is \emph{invalid} if one of the following occurs:
When one of the following occurs, it is easy to see that the 4-tuple does not lead to a required partition.
We say that the 4-tuple $\Psi$ is \emph{invalid} if one of the below cases occur:
\begin{enumerate}[label=(\roman*)]
\item $G[A_i]$ or $G[B_i]$ contains a $C_4$.
\item There exists a pair $\{x,y\}$ $\in$ $P_i$ with a common neighbor in $A_i$.
\item There exists a pair $\{x,y\}$ $\in$ $Q_i$ with a common neighbor in $B_i$.
\end{enumerate}

Note that it is easy to check if a given $\Psi$ is invalid. Below we explain how to compute $M_i[\Psi]$ value at each node $i$.

\vspace{0.1in}
\noindent\textbf{Leaf node:}
For a leaf node $i$, $\Psi = (\emptyset, \emptyset, \emptyset, \emptyset)$ and $M_{i}[\Psi] = 1$.

\vspace{0.1in}
\noindent \textbf{Introduce node:}
Let $j$ be the only child of the node $i$. Suppose $v \in X_i$ is the new vertex present in $X_i$, $v \notin X_j$.
Let $\Psi = (A_i, B_i, P_i, Q_i)$ be a $4$-tuple of $X_i$, If $\Psi$ is invalid, we set $M_i[\Psi]$ to $0$. Otherwise,
we use the following cases to compute the $M_i[\Psi]$ value.

\begin{description}
  \item[Case 1, $v \in A_i$:] If $\exists \{v, x\} \in P_i$ for some $x \in A_i$ or if $\exists \{x,y\} \in P_i$ such that $\{x,y\} \subseteq N(v) \cap A_i$, then $M_i[\Psi] = 0$. Otherwise, $M_i[\Psi] = M_j[\Psi']$,
  where $\Psi' = (A_i \backslash v, B_i, P_i, Q_i)$.

  As $v$ is a newly introduced vertex, it cannot have any forgotten neighbors.
  Hence, $\{v,x\} \in P_i \Longrightarrow M_i[\Psi] = 0$. If $x$ and $y$ have a common forgotten neighbor, they all form a $C_4$, together with $v$. Hence
  $\{x,y\} \in P_i \Longrightarrow M_i[\Psi] = 0$.

\item[Case 2, $v \in B_i$:] If $\exists \{v, x\} \in Q_i$ for some $x \in B_i$ or if $\exists \{x,y\} \in Q_i$ such that $\{x,y\} \subseteq N(v) \cap B_i$, then $M_i[\Psi] = 0$.. Otherwise, $M_i[\Psi] = M_j[\Psi']$,
where $\Psi' = (A_i, B_i \backslash v, P_i, Q_i)$.
\end{description}

\noindent \textbf{Forget node:}
Let $j$ be the only child of the node $i$. Suppose $v \in X_j$ is the vertex missing in $X_i$,
$v \notin X_i$. Let $\Psi = (A_i, B_i, P_i, Q_i)$ be a $4$-tuple of $X_i$, If $\Psi$ is invalid, we set $M_i[\Psi]$ to $0$.
Otherwise, $M_i[\Psi]$ is computed as follows:

%Let $A_j=A_i \cup \{v\}$ and $B_j=B_i \cup \{v\}$.

\begin{description}
  \item[Case 1, $v \in A_j$:] If $\exists x, y \in A_i$ such that $xv, yv \in E$, then $v$ is a
  common
  forgotten neighbor for $x$ and $y$.
  Hence we set  $M_i[\Psi] = 0$ whenever $\{x,y\} \notin P_i$.
  Otherwise, let $R = \{ \{x, y\} | x, y \in A_i \cap N(v)\}$.
  At node $j$, note that any pair in $R$ with a common forgotten neighbor will form a $C_4$. Hence we
   consider
  only those $P_j$'s that are disjoint with $R$.
%  If $x, y \in N(v)$, if $\{x, y\} \in P_j$, we will have a $C_4$, hence the set $R \cap P_j = \emptyset$.
   Also there can be new pairs formed with $v$ at the node $j$. Let $S = \{ \{v, x\} | x \in A_i\}$.
   We have the following equation.
\begin{equation*}\label{c4eq1}
\delta_1 = \max_{X \subseteq S} \{ M_j[A_i \cup v, B_i, (P_i \backslash R) \cup X, Q_i] \}.
\end{equation*}
\item[Case 2, $v \in B_j$:] This is analogous to Case 1.
We set  $M_i[\Psi] = 0$, whenever $\{x,y\} \notin Q_i$.
%If $\exists x, y \in B_i$ such that $xv, yv \in E$, as $x$ and $y$ has common forgot neighbor $v$,
 % $\{x,y\} \in Q_i$, otherwise, we set $M_i[\Psi] = 0$.
  Otherwise, let $R = \{ \{x, y\} | x, y \in B_i \cap N(v) \}$ and  $S = \{ \{v, x\} | x \in B_i\}$.
\begin{equation*}\label{c4eq2}
\delta_2 = \max_{X \subseteq S} \{ M_j[A_i, B_i \cup v, P_i, (Q_i \backslash R) \cup X] \}.
\end{equation*}
\end{description}
If $M_i[\Psi]$ is not set to $0$ already, we set $M_i[\Psi] = \max \{\delta_1, \delta_2 \}$.

\vspace{0.1in}
\noindent \textbf{Join node:}
Let $j_1$ and $j_2$ be the children of the node $i$. By the property of nice tree decomposition, we have
$X_i = X_{j_1} = X_{j_2}$ and $V(T_{j_1}) \cap V(T_{j_2}) = X_i$.
There are no edges between $V(T_{j_1}) \backslash X_i$ and $V(T_{j_2}) \backslash X_i$. Let $\Psi = (A_i, B_i, P_i, Q_i)$
be a $4$-tuple of $X_i$.
%If $G[A_i]$ or $G[B_i]$ has $C_4$ as a subgraph,
If $\Psi$ is invalid, we set $M_i[\Psi]$ to $0$. Otherwise, we use the
following expression to compute the value of $M_i[\Psi]$.

A pair $\{x, y \} \in P_i$ can come either from the left subtree
or from the right subtree but not from both, for that would imply two distinct common neighbors for $x$ and $y$ and hence
a $C_4$. For $X \subseteq P_i$ and $Y \subseteq Q_i$, $\Psi_1 = (A_i, B_i, X, Y)$
and $\Psi_2 = (A_i, B_i, P_i \backslash X, Q_i\backslash Y)$.
\begin{equation*}\label{c4eq3}
  M_i[\Psi] =
  \left\{
    \begin{aligned}[lr]
      &1, & \hbox{$\exists X \subseteq P_i, Y \subseteq Q_i$ such that  $M_{j_{1}}[\Psi_1] = M_{j_{2}}[\Psi_2] = 1$.} \\
      &0, & \hbox{Otherwise.}
    \end{aligned}
  \right.
\end{equation*}

The correctness of the algorithm is implied by the correctness of $M_i[\Psi]$ values, which follows by
a bottom-up induction on the nice tree decomposition.
$G$ has a valid bipartitioning if there exists a 4-tuple $\Psi$ such that
$M_r[\Psi] = 1$, where $r$ is the root of the nice tree decomposition.
%At each node $i$, let $\Delta_i = \max_{\Psi} \{ M_i[\Psi] \}$. If $\Delta_i = 1$, then $G[T_i]$ has a valid partitioning.
%The graph $G$ has a valid bipartitioning if $\Delta_r = 1$, where $r$ is the root node of the nice tree decomposition,

The time complexity at each of the nodes in the tree decomposition is as follows:
constant time at leaf nodes, $O^*(2^{t + t^2})$ time at insert nodes, $O^*(2^{2t + t^2})$ time at forget nodes
and $O^*(2^{t + 2t^2})$ time at join nodes.
This gives the following:
\begin{theorem}
There is an $O(2^{O(t^2)}\dot n)$ time algorithm that solves the \bwhp
when $H = C_4$ on graphs with tree-width at most $t$.
\end{theorem}

\section{Bipartitioning without $H$}
\label{algoH}
\begin{figure}[t]
\begin{minipage}[b]{0.4\textwidth}
\centering
\includegraphics[scale=0.4]{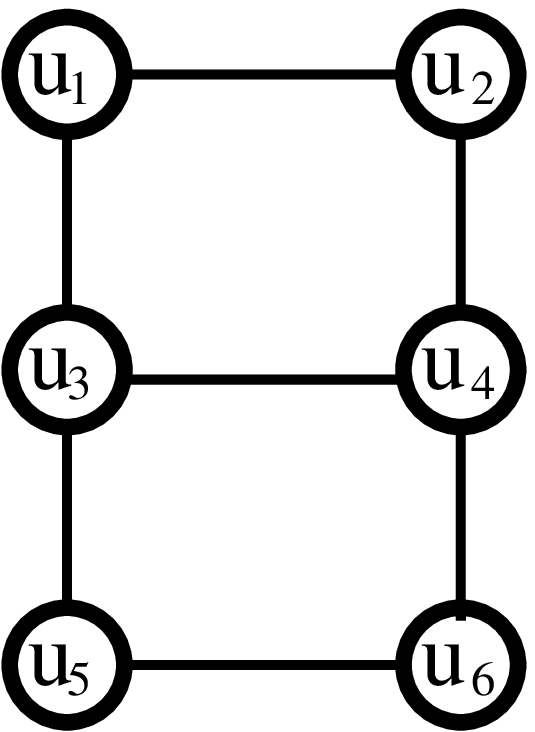}
\caption{An example graph $H$.}
\label{fig:graphH}
\end{minipage}
\hfill
\begin{minipage}[b]{0.5\textwidth}
\centering
\includegraphics[scale=0.4]{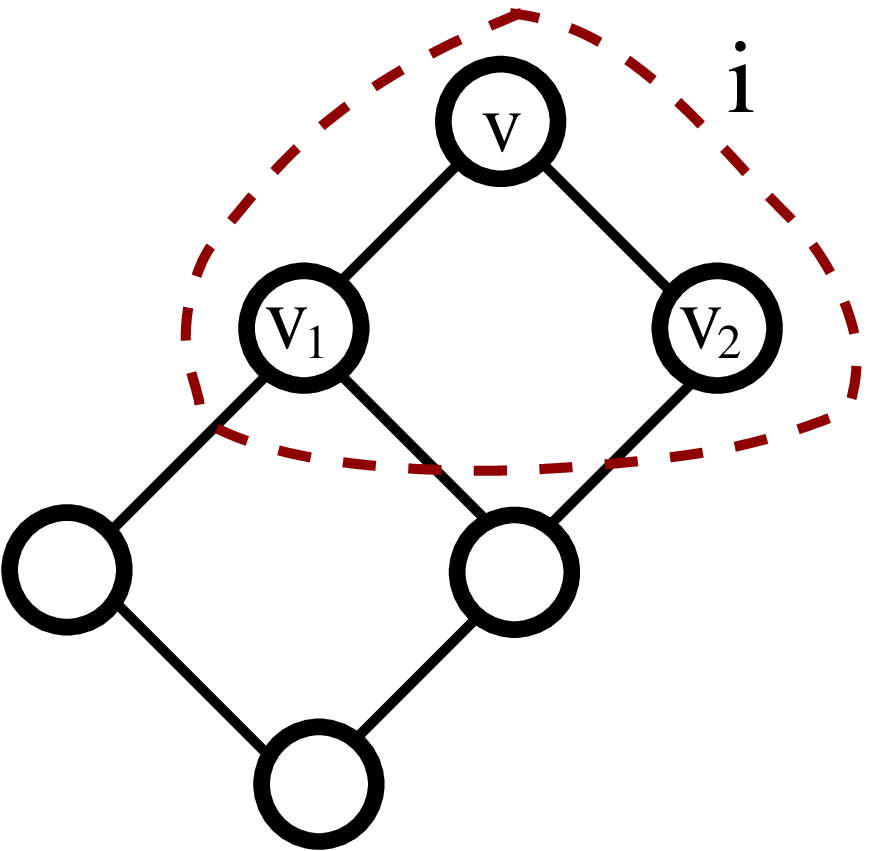}
\caption{Forming $H$ at an introduce node. Sequence $s = (v, v2, v1, fg, fg, fg)$.}
\label{fig:Hinsert}
\end{minipage}
\end{figure}

\begin{figure}[t]
\centering
\includegraphics[scale=0.4]{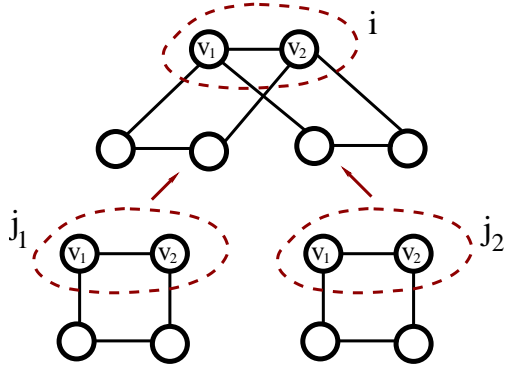}
\caption{Forming $H$ at join node. Sequences at node $j_1$ $s' = (dc, dc, v1, v2, fg, fg)$, at node $j_2$
$s'' = (fg, fg, v1, v2, dc, dc)$ gives a sequence $s = (fg, fg, v1, v2, fg, fg)$ at node $i$.
Vertices outside the dashed lines are forgotten vertices.}
\label{fig:Hjoin}
\end{figure}

Let $X_i$ be a bag at node $i$ of the nice tree decomposition. Let $(A_i, B_i)$ be a partition
of $X_i$. We can easily check if $G[A_i]$ or $G[B_i]$ has $H$ as a subgraph. Otherwise, we need
to see if there is a partition $(A,B)$ of $V(T_i)$ such that $A_i \subseteq A$, $B_i \subseteq B$
and both $G[A_i]$ and $G[B_i]$ do not have $H$ has a subgraph. If there is such a partition $(A,B)$,
then $G[A]$ and $G[B]$ may have subgraph $H'$, an induced subgraph of $H$ which can lead to $H$
at some ancestral node (introduce node or join node) of the nice tree decomposition (See Figures~\ref{fig:Hinsert}
and~\ref{fig:Hjoin}).

We perform dynamic programming over the nice tree decomposition. At each node $i$ we guess a partition
$(A_i,B_i)$ of $X_i$ and possible induced subgraphs of $H$ that are part of $A$ and $B$ respectively.
We check if such a partition is possible. Below we explain the algorithm in detail.

Let the vertices of the graph $H$ are labeled as $u_1, u_2, u_3, \dots, u_r$.
Let $(A_i, B_i)$ be a partition of vertices in the bag $X_i$. Let $(A, B)$ be a partition of $V(T_i)$
such that $A \supseteq A_i$ and $B \supseteq B_i$. We define $\Gamma_{A_i}$ as follows:

\begin{align}
\nonumber
S_{A_i} =& \{(w_1, w_2, w_3, \dots, w_r) | w_\ell \in \{ A_i \cup \{fg, dc\}\}, \\
\nonumber
& \forall \ell_1 \neq \ell_2, w_{\ell_1} = w_{\ell_2} \Longrightarrow w_{\ell_1} \in \{fg, dc\}\}.\\
\nonumber
I_{A_i} =& \{ s = (w_1, w_2, w_3, \dots, w_r) \in S_{A_i} | \mbox{ there exists } \ell_1 \neq \ell_2 \\
\nonumber
& \mbox{ such that } w_{\ell_1} = fg, w_{\ell_2} = dc \mbox{ and } \{u_{\ell_1},u_{\ell_2}\} \in E(H)\}\\
\nonumber
\Gamma_{A_i} =& S_{A_i} \backslash I_{A_i}
\end{align}

Here $fg$ represents a vertex in $A \backslash A_i$, the forgotten vertices in $A$ and $dc$ stands for don't care.
That is we don't care if the corresponding vertex is part of the subgraph or not.
Similarly, we can define $\Gamma_{B_i}$ with respect to the sets $B_i$ and $B$.

A sequence in $S_{A_i}$ corresponds to a subgraph $H'$ of $H$ in $A$ as follows:
\begin{enumerate}
  \item If $w_\ell = fg$ then $u_\ell$ is part of $A \backslash A_i$, the forgotten vertices in $A$.
  \item If $w_\ell = dc$ then $u_\ell$ need not be part of the subgraph $H'$.
  \item If $w_\ell \in A_i$ then the vertex $w_\ell$ corresponds to the vertex $u_\ell$ of $H'$.
\end{enumerate}

$\Gamma_{A_i}$ is the set of sequences that can become $H$ in future at some ancestral (insert/join) node
of the tree decomposition. Note that the sequences $I_{A_i}$ are excluded from $\Gamma_{A_i}$
because a forgot vertex cannot have an edge to a vertex which will come in future at some
ancestral node (insert or join nodes).

\begin{definition}[Subgraph Legal Sequence in $\Gamma_{A_i}$ with respect to $A$]
\label{def1}
A sequence $s = (w_1, w_2, w_3, \dots, w_r) \in \Gamma_{A_i}$ is legal if the sequence $s$ corresponds
to subgraph $H'$ of $H$ within $A$ as follows.
\begin{comment}
\begin{enumerate}
  \item $\forall i, j$ if $w_i = v' \in A_i$ and $w_j = v'' \in A_i$, $\{u_i, u_j\} \in E(H) \Longrightarrow \{v', v''\} \in E(G)$.
  \item $FV(s) = \{ i | w_i = fg\}$, Then there exist $|FV(s)|$ distinct vertices in $A \backslash A_i$ such that:
  \begin{itemize}
    \item $\forall i, j$ if $w_i = fg$, $w_j = v' \in A_i$, $\{u_i, u_j\} \in E(H) \Longrightarrow \{z_i, v'\} \in E(G)$.
    \item $\forall i, j$ if $w_i = w_j = fg$, $\{u_i, u_j\} \in E(H) \Longrightarrow \{z_i, z_j\} \in E(G)$.
  \end{itemize}
\end{enumerate}
\end{comment}

Let $FV(s) = \{ \ell | w_\ell = fg\}$, $DC(s) = \{ \ell | w_\ell = dc\}$ and $VI(s) = [r] \backslash \{FV(s) \cup DC(s)\}$.
Let $H'$ be the induced subgraph of $H$ formed by $u_\ell$, $\ell \in \{VI(s) \cup FV(s)\}$. That is
$H' = H[\{u_\ell| \ell \in VI(s) \cup FV(s)\}]$.

If there exist $|FV(s)|$ distinct vertices $z_\ell \in A \backslash A_i$ corresponding to each index in $FV(s)$ such that
$H'$ is subgraph of  $G[\{ w_\ell | \ell \in VI(s)\} \cup \{z_\ell | \ell \in FV(s)\}]$, then $s$ is legal.
Otherwise, the sequence is illegal.
\end{definition}
Similarly, we define legal/illegal sequences in $\Gamma_{B_i}$ with respect to $B$.

Let $\Psi = (A_i, B_i, P_i, Q_i)$ be a $4$-tuple. Here, $(A_i, B_i)$ is a partition of $X_i$,
$P_i \subseteq \Gamma_{A_i}$ and $Q_i \subseteq \Gamma_{B_i}$.

We define $M_i[\Psi]$ to be $1$ if there is a partition $(A, B)$ of $V(T_i)$ such that:
\begin{enumerate}
  \item $A_i \subseteq A$ and $B_i \subseteq B$.
  \item Every sequence in $P_i$ is legal with respect to $A$.
  \item Every sequence in $Q_i$ is legal with respect to $B$.
  \item Every sequence in $\Gamma_{A_i} \backslash P_i$ is illegal with respect to $A$.
  \item Every sequence in $\Gamma_{B_i} \backslash Q_i$ is illegal with respect to $B$.
  \item Neither $G[A]$ nor $G[B]$ contains $H$ as a subgraph.
\end{enumerate}

Otherwise $M_i[\Psi]$ is set to $0$.

We call a $4$-tuple $\Psi$ as invalid if one of the following conditions occur. If $\Psi$ is invalid
we set $M_i[\Psi]$ to $0$.

\begin{enumerate}
  \item There exists a sequence $s \in P_i$ such that $s$ does not contain $dc$.
  \item There exists a sequence $s \in Q_i$ such that $s$ does not contain $dc$.
\begin{comment}
  \item There exists a sequence $s \in P_i$ which can correspond to the graph $H$ by performing the
  following operation: replace all the $dc$'s in the sequence $s$ with distinct vertices in $A_{i}$ which
  are not already present in $s$.
  \item There exists a sequence $s \in Q_i$ which can correspond to the graph $H$ by performing the
  following operation: replace all the $dc$'s in the sequence $s$ with distinct vertices in $B_{i}$ which
  are not already present in $s$.

  \item There exists a sequence $s = (w_1, w_2, w_3, \dots, w_r) \in \Gamma_{A_i}$ such that the following holds:
  \begin{itemize}
    \item There exist $i \neq j$ such that $w_i = v_{q_1} \in A_i$, $w_j = v_{q_2} \in A_i$.
    \item $\{u_i, u_j\} \in E(H)$.
    \item $\{v_{q_1}, v_{q_2}\} \notin E(G)$.
  \end{itemize}
  \item There exists a sequence $s = (w_1, w_2, w_3, \dots, w_r) \in \Gamma_{B_i}$ such that the following holds:
  \begin{itemize}
    \item There exist $i \neq j$ such that $w_i = v_{q_1} \in B_i$, $w_j = v_{q_2} \in B_i$.
    \item $\{u_i, u_j\} \in E(H)$.
    \item $\{v_{q_1}, v_{q_2}\} \notin E(G)$.
  \end{itemize}
\end{comment}
\end{enumerate}

Now we explain how to compute $M_i[\Psi]$ values at the leaf, introduce, forgot and join nodes of the
nice tree decomposition.

\noindent\textbf{Leaf node:}
Let $i$ be a leaf node, $X_i = \emptyset$, for $\Psi = (A_i, B_i, P_i, Q_i)$, we have $M_i[\Psi] = 1$.
Here $A_i = B_i = \emptyset$, $P_i \subseteq \{([dc]^r)\}$ and $Q_i \subseteq \{([dc]^r)\}$.

\noindent \textbf{Introduce node:}
Let $i$ be an introduce node and $j$ be the child node of $i$. Let $\{v\} = X_i \backslash X_j$.
Let $\Psi = (A_i, B_i, P_i, Q_i)$ be a $4$-tuple at node $i$. If $\Psi$ is invalid we set
$M_i[\Psi] = 0$. Otherwise depending on whether $v \in A_i$ or $v \in B_i$ we have two cases.
We discuss only the case $v \in A_i$, the case $v \in B_i$ can be analogously defined.
\begin{description}
  \item[$v \in A_i$:] We set $M_i[\Psi] = 0$, if there exists an illegal sequence $s$ (in $P_i$)
  containing $v$ or if there exists a trivial legal sequence $s$ containing $v$ but $s$ is not in $P_i$.

  That is, we set $M_i[\Psi] = 0$ in one of the following ($\star$) conditions occurs:
  \begin{framed}
\vspace*{-0.10cm}
  \begin{enumerate}
    \item $\exists \ell_1 \neq \ell_2$, such that $w_{\ell_1} = v$, $w_{\ell_2} \in A_i$, $\{u_{\ell_1}, u_{\ell_2}\} \in E(H)$ but $\{v, w_{\ell_2}\}\notin E(G)$.
    \item $\exists \ell_1 \neq \ell_2$, such that $w_{\ell_1} = v$, $w_{\ell_2} = fg$, $\{u_{\ell_1}, u_{\ell_2}\} \in E(H)$.
     \item Let $s = (w_1, w_2, w_3, \dots, w_r) \in \Gamma_{A_i} \backslash P_i$.
    There exists $\ell_1$ such that $w_{\ell_1} = v$ and for all $\ell_2 \neq \ell_1$ $w_{\ell_2} \in A_i \cup \{dc\}$.
  For all $\ell_1 \neq \ell_2$ $w_{\ell_1}, w_{\ell_2} \in A_i$, $\{u_{\ell_1}, u_{\ell_2}\} \in E(H)$ $\Longrightarrow$ $\{w_{\ell_1}, w_{\ell_2}\} \in E(G)$.
  \end{enumerate}
 \vspace*{-0.10cm}
\end{framed}
  Otherwise we set $M_i[\Psi] = M_j[\Psi']$, where $\Psi' = (A_i \backslash v, B_i, P_j, Q_i)$. Here $P_j$
  is computed as follows:
 \begin{definition}
  $\mbox{Rep}_{dc}(s, v) = s'$, sequence $s'$ obtained by replacing $v$ (if present) with $dc$ in $s$.
 \end{definition}
 Note that, $\mbox{Rep}_{dc}(s, v) = s$, if $v$ not present in $s$.

$P_j = \cup_{s \in P_i} \{\mbox{Rep}_{dc}(s, v)\}$.
\end{description}
\noindent \textbf{Forget node:}
Let $i$ be a forget node and $j$ be the only child of node $i$. Let $\{v\} = X_j \backslash X_i$.
Let $\Psi = (A_i, B_i, P_i, Q_i)$ be a $4$-tuple at node $i$. If $\Psi$ is
invalid we set $M_i[\Psi] = 0$. Otherwise, we set $M_i[\Psi] = \max \{\delta_1, \delta_2 \}$ where
$\delta_1$ and $\delta_2$ are computed as follows:
\begin{description}
  \item[Computing $\delta_1$:] Set $A_j = A_i \cup \{v\}$. As $v$ is the extra vertex in $A_j$, there
  could be many possible $P_j$ at node $j$.
  \begin{definition}
  $\mbox{Rep}_{fg}(s, v) = s'$, sequence $s'$ obtained by replacing $v$ (if present) with $fg$ in $s$.
  \end{definition}
  Note that, if $s$ does not contain the vertex $v$ then $\mbox{Rep}_{fg}(s, v) = s$.

  We also extend the definition of $\mbox{Rep}_{fg}$ to a set of sequences as follows:
  \begin{equation}
    \nonumber
    \mbox{Rep}_{fg}(S, v) = \cup_{s \in S}\{ \mbox{Rep}_{fg}(s, v) \}.
  \end{equation}
 Note that, if $s$ is a legal sequence at the node $j$ with respect to $A$, then $\mbox{Rep}_{fg}(s, v)$
 is also a legal sequence at node $i$ with respect to $A$.

   $$\delta_1 = \max_{\substack{P_j \subseteq \Gamma_{A_j} \\ \mbox{\scriptsize Rep}_{fg}(P_j, v) = P_i}} \{M_j[(A_j, B_i, P_j, Q_i)]\}$$

  \item[Computing $\delta_2$:] $B_j = B_i \cup v$. It is analogous to computing $\delta_1$ but we process on
  $B$.
\end{description}
\noindent \textbf{Join node:}
Let $i$ be a join node, $j_1$, $j_2$ be the left and right children of the node $i$ respectively.
$X_i = X_{j_1} = X_{j_2}$ and there are no edges between $V(T_{j_1})\backslash X_i$ and $V(T_{j_2})\backslash X_i$.
Let $\Psi = (A_i, B_i, P_i, Q_i)$ be a $4$-tuple at node $i$. If $\Psi$ is invalid we set $M_i[\Psi] = 0$.
Otherwise, we compute $M_i[\Psi]$ value as follows:
\begin{definition}
Let $s = (w_1, w_2, w_3, \dots, w_r)$, $s' = (w_1', w_2', w_3', \dots, w_r')$ and $s'' = (w_1'', w_2'', w_3'', \dots, w_r'')$
be three sequences. We say that $s = \mbox{Merge}(s', s'')$ if the following conditions are satisfied.
\begin{enumerate}
  \item $\forall \ell$ $w_\ell \in X_i$  $\Longrightarrow$ $w_\ell' = w_\ell'' = w_\ell$.
  \item $\forall \ell$ $w_\ell = fg$ $\Longrightarrow$ either $(w_\ell' = fg \mbox{ and } w_\ell'' = dc)$ or
  $(w_\ell' = dc \mbox{ and } w_\ell'' = fg)$.
  \item $\forall \ell$ $w_\ell = dc$ $\Longrightarrow$ $w_\ell' = w_\ell'' = dc$.
\end{enumerate}
\end{definition}
Note that, if $s' \in \Gamma_{A_{j_1}}$ and $s'' \in \Gamma_{A_{j_2}}$ are legal sequences at node $j_1$
and $j_2$ respectively then $s$ is a legal sequence at node $i$ with respect to $A$.
We extend the Merge operation to sets of sequences as follows:
$$\mbox{Merge}(S_1, S_2) = \{s | \exists s' \in S_1, s'' \in S_2 \mbox{ such that } s = \mbox{Merge}(s', s'')\}.$$

We set $M_i[\Psi] = 1$ if there exists $P_{j_1}$, $Q_{j_1}$, $P_{j_2}$ and $Q_{j_2}$  such that
the following conditions are satisfied:

\begin{tabular}{ll}
(i) $P_i = \mbox{Merge}(P_{j_1}, P_{j_2})$,
&(ii) $Q_i = \mbox{Merge}(Q_{j_1}, Q_{j_2}),$\\
(iii) $M_{j_1}[A_i, B_i, P_{j_1}, Q_{j_1}] = 1$, and \hspace{0.1in}
&(iv) $M_{j_2}[A_i, B_i, P_{j_2}, Q_{j_2}] = 1$.
\end{tabular}

The graph has valid bipartitioning if there exists a $\Psi$ such that $M_r[\Psi] = 1$.
Where $r$ is the root node of the nice tree decomposition. The correctness of the algorithm
is implied by the correctness of $M_i[\Psi]$ values, which can be proved using a bottom up
induction on the nice tree decomposition. The time complexity of the algorithm is $O^*(2^{2t^r})$.
Thus we get the following:
\begin{theorem}
\label{HS:thm1}
There is an $2^{O(t^r)} \cdot n$ time algorithm that solves the \bwh problem for any arbitrary fixed
$H$ ($|V(H)| = r$), on graphs with tree-width at most $t$.
\end{theorem}

\section{$H$-Free Bipartitioning Problem}
\label{sec:hfree}
The techniques described in Section~\ref{algoH} can also be used to solve the \hfbp. As we are looking
for bipartitioning without $H$ as an induced subgraph. Definition~\ref{def1} and ($\star$) conditions at
the introduced node are modified as below.
\begin{definition}[Induced Subgraph Legal Sequence in $\Gamma_{A_i}$ with respect to $A$]
A sequence $s = (w_1, w_2, w_3, \dots, w_r) \in \Gamma_{A_i}$ is legal if the sequence $s$ corresponds
to subgraph $H'$ of $H$ within $A$ as follows.

Let $FV(s) = \{ \ell | w_\ell = fg\}$, $DC(s) = \{ \ell | w_\ell = dc\}$ and $VI(s) = [r] \backslash \{FV(s) \cup DC(s)\}$.
Let $H'$ be the induced subgraph of $H$ formed by $u_\ell$, $\ell \in \{VI(s) \cup FV(s)\}$. That is
$H' = H[\{u_\ell| \ell \in VI(s) \cup FV(s)\}]$.

If there exist $|FV(s)|$ distinct vertices $z_\ell \in A \backslash A_i$ corresponding to each index in $FV(s)$ such that
$H'$ is isomorphic to  $G[\{ w_\ell | \ell \in VI(s)\} \cup \{z_\ell | \ell \in FV(s)\}]$, then $s$ is legal.
Otherwise, the sequence is illegal.
\end{definition}

($\star$) conditions at the introduced node:
\begin{framed}
\vspace*{-0.50cm}
  \begin{enumerate}
    \item $\exists \ell_1 \neq \ell_2$, such that $w_{\ell_1} = v$, $w_{\ell_2} \in A_i$, $\{u_{\ell_1}, u_{\ell_2}\} \in E(H)$ but $\{v, w_{\ell_2}\}\notin E(G)$.
    \item $\exists \ell_1 \neq \ell_2$, such that $w_{\ell_1} = v$, $w_{\ell_2} \in A_i$, $\{u_{\ell_1}, u_{\ell_2}\} \notin E(H)$ but $\{v, w_{\ell_2}\}\in E(G)$.
    \item $\exists \ell_1 \neq \ell_2$, such that $w_{\ell_1} = v$, $w_{\ell_2} = fg$, $\{u_{\ell_1}, u_{\ell_2}\} \in E(H)$.
     \item Let $s = (w_1, w_2, w_3, \dots, w_r) \in \Gamma_{A_i} \backslash P_i$.
    There exists $\ell_1$ such that $w_{\ell_1} = v$ and for all $\ell_2 \neq \ell_1$ $w_{\ell_2} \in A_i \cup \{dc\}$.
  For all $\ell_1 \neq \ell_2$ $w_{\ell_1}, w_{\ell_2} \in A_i$, $\{u_{\ell_1}, u_{\ell_2}\} \in E(H)$ $\Longleftrightarrow$ $\{w_{\ell_1}, w_{\ell_2}\} \in E(G)$.
  \end{enumerate}
 \vspace*{-0.50cm}
\end{framed}
Thus we get the following:
\begin{theorem}
\label{HF:thm2}
There is an $2^{O(t^r)} \cdot n$ time algorithm that solves the \hfbp
 for any arbitrary fixed $H$ ($|V(H)| = r$), on graphs with tree-width at most $t$.
\end{theorem}

\noindent \textbf{Coloring without subgraph $H$:}
We note that our techniques extend in a straightforward manner to solve 
the $q$-coloring analogues of \bwh and \hfb problems.
where we have to partition 
the vertices of $G$ into $q$ sets such that graphs induced by none of 
these sets have $H$ as a subgraph or induced subgraph.
In this case, we have to consider tuples $\Psi$ that have $2q$ sets. 
%\\That is $\Psi = (A_i^1, A_i^2, \dots, A_i^q, P_i^1, P_i^2, \dots, P_i^q)$.
%Here $A_i^j \subseteq X_i$ and $P_i^j \subseteq \Gamma_{A_i^j}$. 
The operations at the leaf, introduce and forget nodes
are very similar to the case of bipartitioning. At the join node we need to define the
Merge operation on $q$ sets instead of $2$ sets. The running time of these algorithms 
are similar to that of the algorithms that solve the bipartitioning problems.

We further consider the optimization problems of finding the smallest $q$ for which 
$V(G)$ can be partitioned into $q$ sets such that graphs induced by none of 
these sets have $H$ as a subgraph or an induced subgraph. Since the chromatic 
number of $G$ is at most $t + 1$ (where $t$ is the tree-width of $G$), 
the algorithm needs to search for the smallest $q \leq t+1$.
%%WE NEED citation for the above.
Thus we get the following:

\begin{theorem}
The problem of finding the smallest $q$ for which $V(G)$ can be partitioned into $q$ 
sets such that the graphs induced by none of these parts have $H$ as a subgraph (or as an 
induced subgraph) is FPT when parameterized by tree-width.
\end{theorem}

%the $q$-coloring problems will be esse
%
%Thus we have the following:
%\begin{theorem}
%\label{HF:thm3}
%There is an $q^{O(t^r)} \cdot n$ time algorithm that solves the \hfqcp
% for any arbitrary fixed $H$ ($|V(H)| = r$), on graphs with tree-width at most $t$.
%\end{theorem}

%\section{Conclusions}
%In this paper we consider the \bwhp, where we study if it is possible to bipartition a graph into two sets
%such that neither set has graph $H$ as a subgraph. We present combinatorial FPT algorithms for fixed
%graph $H$. The techniques can also be used to solve \hfbp and \hfqcp.
\bibliographystyle{splncs}
\bibliography{BibFile}
\end{document}